\documentclass[print]{aa}  
\usepackage[switch]{lineno}
\usepackage{graphicx}
\usepackage{txfonts}
\usepackage{lipsum}
\usepackage{subcaption}        
                               
\usepackage{lscape}            
                               
\usepackage{placeins}          
                               
\usepackage{amsmath}
\usepackage{multirow}
\usepackage{caption}
\usepackage{xcolor}

\newcommand{\kmspc}{${\rm km\ s^{-1} Mpc^{-1}}$}
\newcommand{\pccm}{${\rm pc\ cm^{-3}}$}
\newcommand{\DMIGM}{${\rm DM_{IGM}}$}
\newcommand{\DMhost}{${\rm DM_{host}}$}
\newcommand{\DMISM}{${\rm DM_{ISM}}$}
\newcommand{\DMhalo}{${\rm DM_{halo}}$}
\newcommand{\DMexc}{${\rm DM_{exc}}$}
\newcommand{\DMobs}{${\rm DM_{obs}}$}

\newcommand{\emu}{$e^{\mu}$}

\begin{document}

   \title{Measuring the Hubble constant using localized and nonlocalized fast radio bursts}
	\titlerunning{H$_0$ from fast radio bursts}
	
   \author{D. H. Gao\inst{1}
        \and Q. Wu\inst{1}
        \and J. P. Hu\inst{1}
        \and S. X. Yi\inst{2}
        \and X. Zhou\inst{3}$^,$\inst{4}
        \and F. Y. Wang\inst{1}$^,$\inst{5}
        \fnmsep\thanks{E-mail: fayinwang@nju.edu.cn}
        \and Z. G. Dai\inst{6}
        \fnmsep\thanks{E-mail: daizg@ustc.edu.cn}
        }\authorrunning{Gao et al. }

   \institute{School of Astronomy and Space Science, Nanjing University, Nanjing 210093, China\\
            \and School of Physics and Physical Engineering, Qufu Normal University, Qufu 273165, China\\
            \and Xinjiang Astronomical Observatory, Chinese Academy of Sciences, Urumqi 830011, China\\
            \and Xinjiang Key Laboratory of Radio Astrophysics, 150 Science1-Street, Urumqi 830011, China\\
            \and Key Laboratory of Modern Astronomy and Astrophysics (Nanjing University), Ministry of Education, Nanjing 210093, China\\
            \and Department of Astronomy, University of Science and Technology of China, Hefei 230026, China}

   \date{Received April 30, 2025}

  \abstract
    {The Hubble constant ($H_0$) is one of the most important parameters in the standard $\rm \Lambda CDM$ model. The measurements given by the main two methods show a gap larger than $4\sigma$, which is known as Hubble tension. Fast radio bursts (FRBs) are extragalactic pulses with durations of milliseconds. They can be used as cosmological probes. We constrain $H_0$ using localized and nonlocalized FRBs. We first used 108 localized FRBs to constrain $H_0$ using the probability distributions of \DMhost and \DMIGM from the IllustrisTNG simulation. Then, we used a Monte Carlo sampling to calculate the pseudo-redshift distributions of 527 nonlocalized FRBs from CHIME observations. The 108 localized FRBs yield a constraint of $H_0=69.40_{-1.97}^{+2.14}$ \kmspc, which lies between the early- and late-time values. The constraint of $H_{0}$ from nonlocalized FRBs yields $H_0=68.81_{-0.68}^{+0.68}$ \kmspc. This result indicates that the uncertainty on the constraint of $H_0$ drops to $\sim1\%$ when the number of localized FRBs is increased to $\sim500$. These uncertainties only include the statistical error. The systematic errors are also discussed and play a dominant role in the current sample.}


   \maketitle

\section{Introduction}
\label{sec:Introduction}
The lambda cold dark matter ($\rm \Lambda CDM$) model has provided a convincing explanation for numerous cosmological observational facts. 
As one of the most fundamental parameters in cosmology, the Hubble constant ($H_0$) describes the expansion rate of the current Universe \citep{Intro-HubbleLaw}, and its reciprocal 1/$H_0$ gives an estimate of the age of the Universe. Constraints on $H_0$ were generally made with two distinct methods \citep{Intro-HubbleReview}: early-time probes given by the cosmic microwave background (CMB), and late-time probes given by stars such as Cepheid-calibrated type Ia supernovae (SNe Ia). With rapidly developing telescopes, the predictions given by the two methods have shown an increased accuracy. \citet{Intro-Planck2018} predicted $H_0 = 67.66\pm0.42$ \kmspc at a confidence level of 68\% based on power spectra from the \textit{Planck} CMB, while \citet{Intro-SNIa} showed $H_0 = 73.04\pm1.04$ \kmspc based on a Cepheid-SNe Ia sample. A non-negligible gap of more than 4$\sigma$ appears between the two results, which is known as the Hubble tension \citep{Intro-HubbleTension, Intro-HubbleTension2}. It is crucial to find an independent approach to resolve the Hubble tension.

Fast radio bursts (FRBs) are extraordinarily bright radio bursts that were first discovered in 2007 \citep{Intro-FirstFRB}. 
With subsequent discoveries of five FRBs in several years \citep{Intro-FRB2, Intro-FRB3}, FRBs are universally recognized as a special type of high-energy astronomical phenomenon. 
They are generally characterized by an extremely high burst energy and a duration time of some milliseconds \citep{Intro-FRBReview2,Petroff2022,Intro-FRBReview,Wu2024}. Almost all FRBs are produced outside the Milky Way because of the extraordinary burst rate and extragalactic dispersion measures \citep{Intro-Extragalactic}. Some FRBs have been observed to repeat, while others have not shown repetitiveness so far.  A small but growing proportion of these FRBs are localized.

To employ FRBs as cosmological probes, the dispersion measure (DM) is a characteristic quantity. It is defined as the integral of the electron number density along the path of propagation, that is, DM=$\int_0^d n_e(l)$d$l$. 
FRBs were used as independent cosmological probes in multiple cases \citep{Bhandari2021,Wu2024}, such as for measuring $H_0$ \citep{Wu2022, Hagstotz2022MNRAS, James2022,Wei2023,Intro-HubbleTaylor,kalita2024} and Hubble parameter \citep{Wu2020}, and finding missing baryons \citep{Macquart2020Natur,Yang2022ApJL,Lin2023MNRAS,Wang2023, connor2024}. 
Early researches assumed that DMs contributed by the intergalactic medium (\DMIGM) and the host galaxy (\DMhost) had certain values, although it is practically not possible to distinguish the partition between \DMIGM and \DMhost. \citet{Wu2022} provided a possible solution to this degeneracy problem by considering the probability density distributions for \DMIGM and \DMhost. Some works also used similar methods to solve the degeneracy problem, such as the FLIMFLAM\footnote{FRB Line-of-sight Ionization Measurement From Lightcone AAOmega Mapping} survey and the $\rm F^4$\footnote{Fast and Fortunate for FRB Follow-up} team \citep{Simha2021ApJ, Lee2022ApJ, Khrykin2024ApJ, Huang2025}.

In this paper, we constrain $H_0$ with 108 localized FRBs and 527 nonlocalized FRBs. In Section~\ref{sec:DM},we introduce the theoretical model we used for the DMs of FRBs. In Section~\ref{sec:localized} we show the Markov chain Monte Carlo method and constrain $H_0$ using localized FRBs. In Section~\ref{sec:unlocalized} we propose a method for constraining $H_0$ using nonlocalized FRBs. In Section~\ref{sec:discussion} we discuss the statistical and systematic errors.

\section{Distribution of the DMs}
\label{sec:DM}
The observed DMs of FRBs were separated into the following components:
\begin{equation}
    {\rm DM_{obs} = DM_{ISM} + DM_{halo} + DM_{IGM}} + \frac{{\rm DM_{host}}}{1+z},
        \label{eq:DM_components}
\end{equation}
where \DMobs is the total observed DM, and \DMISM, \DMhalo, \DMIGM, and \DMhost refer to the DMs that contributed by the interstellar medium (ISM) within the Milky Way, the Galactic halo, the intergalactic medium (IGM), and the host galaxy, respectively. The former two components, \DMISM and \DMhalo, are contributed by the medium within the Milky Way and are often referred to as a whole, that is, $\rm DM_{MW} = DM_{ISM} + DM_{halo}$. \DMISM is well described by Galactic electron distribution models, such as YMW16 by \citet{YMW16} and NE2001 by \citet{NE2001}. $\rm DM_{MW}$ and its uncertainty were well modeled previously \citep{DM_halo,Keating2020MNRAS,Ravi2023-2}. We applied NE2001 to estimate \DMISM. We assumed that \DMhalo ~follows a Gaussian distribution with $\langle \rm DM_{halo} \rangle$ = 65 \pccm and $\sigma=15$ \pccm.

The extragalactic component was obtained by subtracting $\rm DM_{MW}$ from the observed DM,
\begin{equation}
    {\rm DM_{exc} = DM_{obs} - DM_{MW} = DM_{IGM}} + \frac{{\rm DM_{host}}}{1+z}.
        \label{eq:DM_excess}
\end{equation}
In the standard $\rm \Lambda CDM$ universe model, the mean value of \DMIGM~ is 
\begin{equation}
\left\langle{\rm DM_{IGM}}\right\rangle= \frac{3 c H_{0} \Omega_{b} f_{\rm IGM}}{8 \pi G m_{p}} \times f_e(z),
\label{eq:DM_IGM_mean}
\end{equation}
where $m_p$ is the proton mass, and $f_{\rm IGM}$ is the fraction of baryons in the IGM. A value of $f_{\rm IGM}\simeq 0.84$ was preferred previously according to \citet{fIGM}. \citet{connor2024} found a more accurate constraint on $f_{\rm IGM}\simeq 0.93$ with data from the Deep Synoptic Array, however. We adopted $f_{\rm IGM}=0.93$ in our calculation. The integral $f_e(z)$ is defined by
\begin{equation}
f_e(z) = \int_{0}^{z} \frac{\left[\frac{3}{4} y_{1} \chi_{e, {\rm H}}(z)+\frac{1}{8} y_{2} \chi_{e, {\rm He}}(z)\right](1+z) d z}{\left[\Omega_{m}(1+z)^{3}+\Omega_{\Lambda}\right]^{1 / 2}}.
\label{eq:He(z)}
\end{equation}
The cosmological parameters $\Omega_m$ and $\Omega_{\Lambda}$ are given by \citet{Intro-Planck2018}. $\Omega_b$ was determined as an assumption in the Planck results based on Big Bang nucleosynthesis (BBN) constraints and on the primordial deuterium abundance measurements by \citet{Omegabh2}. It is always given in the form of $\Omega_bh^2$, where $h=H_0/100$ \kmspc. We therefore modified the equation to keep $\Omega_b$ in the form of $\Omega_b{H_0}^2$. $y_1$ and $y_2$ in Equation~(\ref{eq:He(z)}) are the hydrogen and helium fractions normalized to 0.75 and 0.25, respectively, which can be neglected as $y_1\simeq y_2\simeq 1$. $\chi_{e, {\rm H}}(z)$ and $\chi_{e, {\rm He}}(z)$ are the ionization fractions of hydrogen and helium, which can also be considered to be $\chi_{e, {\rm H}}(z) = \chi_{e, {\rm He}}(z) = 1$ at $z<3$. Equations~(\ref{eq:DM_IGM_mean}) and~(\ref{eq:He(z)}) are further rewritten as
\begin{equation}
\left\langle{\rm DM_{IGM}}\right\rangle= \frac{21 c  \Omega_{b} {H_0}^2 }{64 \pi H_0 G m_{p}} \times \int_{0}^{z} \frac{f_{\rm IGM} (1+z) d z}{\left[\Omega_{m}(1+z)^{3} + 1 - \Omega_m \right]^{1 / 2}}.
\label{eq:DM_IGM_mean_2}
\end{equation}

\section{Constraint on $H_0$ with localized FRBs}
\label{sec:localized}
\subsection{Monte Carlo sampling}
\label{sec:MCMC}

\begin{figure}
	\includegraphics[width=\columnwidth]{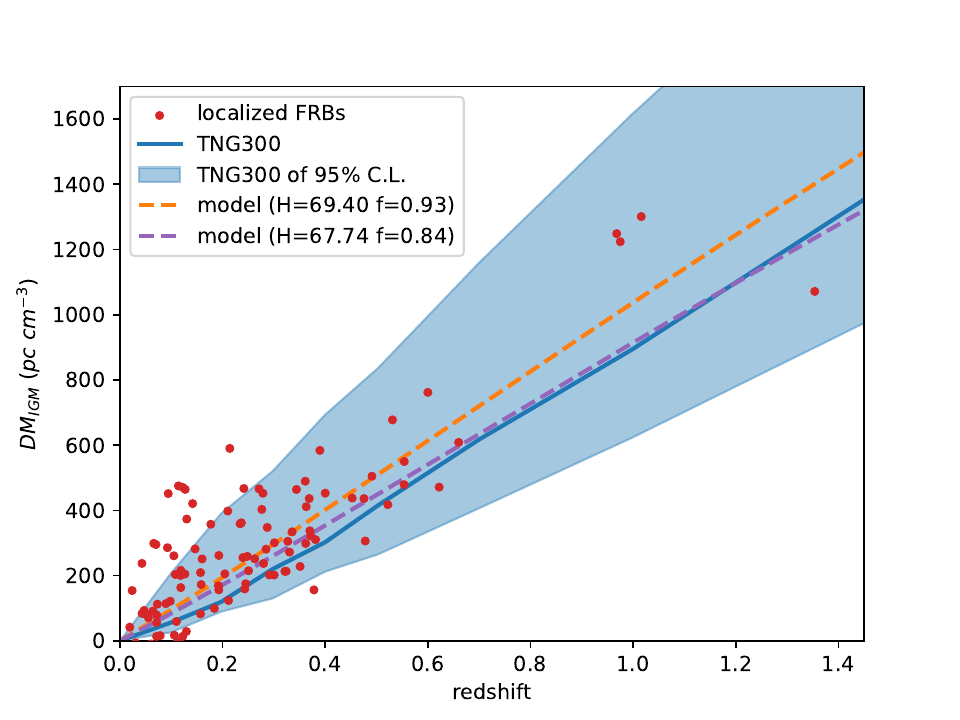}
    \centering
    \caption{\DMIGM-$z$ relation for localized FRBs. The red dots show localized FRBs. The blue line corresponds to the \DMIGM~ from the IllustrisTNG 300 simulation, and the blue shaded area is the 95\% confidence region. The dashed orange and purple lines represent the models in Equation~(\ref{eq:DM_IGM_mean_2}) with different parameters ($H_0$ and $f_{IGM}$).  $\langle$\DMIGM$\rangle$=\DMexc-$\langle$\DMhost$\rangle$, where $\langle$\DMhost$\rangle$ is given by  theIllustrisTNG simulation.}
    \label{fig:model}
\end{figure}

The host galaxies of a few FRBs have been determined. We collected data of all localized FRBs until March 2025, and they are shown in Table~\ref{tab:localized_FRBs} and Fig.~\ref{fig:model}. We list the equatorial coordinates, DMs, and redshifts of the FRBs. All FRBs were classified into three types based on the properties of the host galaxies, which is necessary for the \DMhost~ probability distributions derived from the IllustrisTNG simulation. It should be noted that \DMhost~ can be further divided into two components, which are contributed by its host galaxy and the local environment near the source ($\rm DM_{source}$). FRBs such as FRB20190520B and FRB20220831A have extremely high $\rm DM_{source}$ based on observational data \citep{Wu2022,connor2024}. FRB20190520B is also influenced by the strong DM from intervening galaxies \citep{Lee2023ApJL}. FRBs such as FRB20181030, FRB20200120E, FRB20220319D, and FRB20210405I were excluded because ${\rm DM_{MW}}$ is large, which causes \DMexc$<0$.  FRB20221027A was excluded for its ambiguity in the host galaxy localization \citep{Sharma2024}.

To run a Markov chain Monte Carlo (MCMC) simulation, we calculated the probability distribution of extragalactic DM components. \DMhost ~follows a lognormal distribution \citep{Macquart2020Natur,Zhang2020host}
\begin{equation}
p_{\rm host}\left({\rm DM_{host}}\right)=\frac{1}{\sqrt{2\pi}\ {\rm DM_{host}}\ \sigma_{\rm host}} \exp \left[-\frac{(\ln{\rm DM_{host}}-\mu)^{2}}{2 \sigma_{\rm host}^{2}}\right],
\label{eq:DM_host_pdf}
\end{equation}
where \emu ~ is the mean of the distribution.
\DMIGM ~can be fit with a Gaussian-like distribution, which is written as \citep{Macquart2020Natur,Zhang2021IGM}
\begin{equation}
p_{\rm IGM}(\Delta)=A \Delta^{-\beta} \exp \left[-\frac{\left(\Delta^{-\alpha}-C_{0}\right)^{2}}{2 \alpha^{2} \sigma_{\rm IGM}^{2}}\right],\ \Delta=\frac{\rm DM_{IGM}}{\langle{\rm DM_{IGM}}\rangle},
\label{eq:DM_IGM_pdf}
\end{equation}
where $\alpha=\beta=3$ \citep{Macquart2020Natur}. $A$ and $C_0$ are the normalization parameters given by $\int p_{\rm IGM} = 1$ and $\langle\Delta\rangle=1$.

Previous works were made with similar MCMC methods \citep{James2022,kalita2024}. They fit all free parameters simultaneously. The probability function is
$p\sim p_{\rm host}({\rm DM}\mid e^\mu, \sigma_{\rm host}, H_0)\ p_{\rm IGM}({\rm DM}\mid \sigma_{\rm IGM}, H_0)$.
Since the number of localized FRBs is small ($n_{\rm local}\sim100$ currently, and $n_{\rm local}<20$ for earlier researches), the confidence is significantly weakened to fit four parameters $(e^\mu, \sigma_{\rm host}, \sigma_{\rm IGM}, H_0)$ simultaneously. Previous works also assumed that the distribution parameters were fixed constants for different FRBs, but some showed a dependence on redshift.

To reduce the size of the parameter space, we used the probability distributions derived from the IllustrisTNG simulation. \citet{Zhang2020host} and \citet{Zhang2021IGM} derived the best-fit distribution parameters of \DMhost and \DMIGM from the IllustrisTNG simulation \citep{IllustrisTNG}.  To apply the results from IllustrisTNG simulation, all localized FRBs were divided roughly into three categories based on the properties of the host galaxies: Nonrepeating bursts, FRB121102-like repeating bursts, and FRB180916-like repeating bursts.  \citet{Zhang2020host} and \citet{Zhang2021IGM} provided the best-fit values of the distribution parameters $(e^\mu, \sigma_{\rm host}, \sigma_{\rm IGM}, A, and C_0)$ in Equations (6) and (7) at several redshifts. We performed a monotone cubic spline interpolation for each parameter to obtain the values at any given redshift.

Taking all parameters into Equations~(\ref{eq:DM_host_pdf}) and ~(\ref{eq:DM_IGM_pdf}), we obtained the likelihood function for one FRB,
\begin{equation}
\begin{aligned}
\mathcal{L}_{\rm FRB} = &\int_0^{(1+z){\rm (DM_{obs}-DM_{MW})}} p_{\rm host}({\rm DM_{host}} \mid H_0)\ \times\\
&p_{\rm IGM}({{\rm DM_{obs}-DM_{MW}} - \frac{\rm DM_{host}}{(1+z)}}\mid H_0)\ d\ {\rm DM_{host}}.
\label{eq:likelihood_FRB}
\end{aligned}
\end{equation}
The parameters of the \DMhost~ and \DMIGM ~distributions in Equation~(\ref{eq:likelihood_FRB}) are different for each FRB. For the $i^{\rm th}$ FRB, the complete likelihood function is
\begin{equation}
\mathcal{L}_i=\mathcal{L}_i({{\rm DM}_{\rm obs}^{(i)}}\mid H_0, {{\rm DM}_{\rm MW}^{(i)}}, z_i, e^\mu_i, \sigma_{\rm host}^{(i)}, \sigma_{\rm IGM}^{(i)}, A_i, C_{0,i}).
\label{eq:likelihood_parameters}
\end{equation}
The total log-likelihood function of all FRBs is
\begin{equation}
\begin{aligned}
\ln\mathcal{L}(H_0) = \sum_{i=1}^{n} \ln\mathcal{L}_i({{\rm DM}_{\rm obs}^{(i)}} \mid H_0).
\label{eq:loglikelihood_total}
\end{aligned}
\end{equation}
We used a uniform distribution $H_0\in \mathcal{U}(0,100)$ \kmspc as prior.

\subsection{MCMC sampling of localized FRBs}
\subsubsection{Data preprocessing}

To run the MCMC sampling, we used the open-source Python package \texttt{emcee} \citep{emcee}. We adopted $f_{\rm IGM}=0.93, {\rm DM}_{\rm halo}=65$ \pccm and the cosmological parameters given by \citet{Intro-Planck2018}. The statistical error of these parameters is discussed in Section~\ref{sec:discussion}. Before the initialization of MCMC, the observation data were preprocessed. The preprocessing included the steps listed below.

(a) We calculated the Galactic component of DMs and subtracted it from the total DM to obtain \DMexc.

(b) We performed monotone cubic spline interpolations on the data from \citet{Zhang2020host,Zhang2021IGM} and calculated the distribution parameters involved in Equation~(\ref{eq:likelihood_parameters}) for each FRB.

(c) We set the initial positions for the MCMC walkers. A universal choice for the initialization is to uniformly scatter walkers in a small sphere around the optimal value given by a maximum likelihood estimation (MLE). We tested intervals with a length of $10^{-3}$ \kmspc centered at different values within $[60,80]$ \kmspc, and we found that the initialization has little influence on the constraint result and converged within 30 MCMC steps. We used $\mathcal{U}(70-10^{-3},70+10^{-3})$ \kmspc as our final initialization.

\begin{figure}
	\includegraphics[width=\columnwidth]{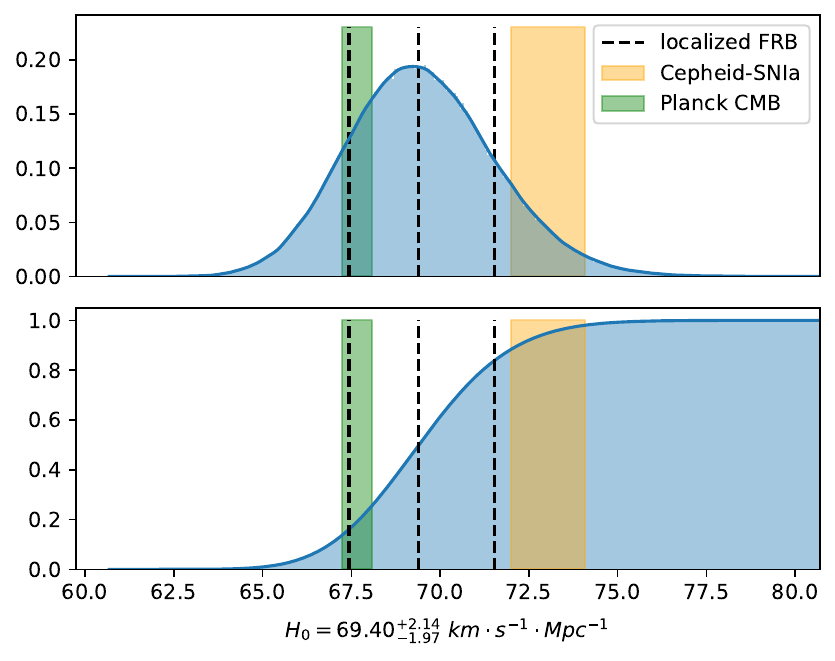}
    \centering
    \caption{Probability density function and cumulative distribution function of $H_0$ given by 108 localized FRBs. The vertical line shows our result $H_0=69.40_{-1.97}^{+2.14}$ \kmspc with 1$\sigma$ uncertainty. 
    	The orange and green regions show the 1$\sigma$ confidence intervals given by SNe Ia and the CMB, respectively.}
    \label{fig:H_0_localized}
\end{figure}
\subsubsection{Monte Carlo cycle and postprocessing}
\label{sec:MCMC_local}
We set up a Monte Carlo system with 512 walkers. In each Monte Carlo cycle, the program went through the steps listed below.

(a) We calculated the mean value of \DMIGM with the current $H_0$ based on Equation~(\ref{eq:DM_IGM_mean_2}) for each FRB.

(b) For any given \DMhost, we calculated $p_{\rm host}({\rm DM_{host}})$ and $p_{\rm IGM}(\Delta)$ based on Equations~(\ref{eq:DM_host_pdf}) and (\ref{eq:DM_IGM_pdf}), and integrated \DMhost ~to obtain the 
likelihood function $\mathcal{L}_{\rm FRB}$ according to Equation~(\ref{eq:likelihood_FRB}) for each FRB.

(c) We summed the log-likelihood functions of all FRBs and updated $H_0$ based on the total likelihood.

The autocorrelation time $\tau_f$ is a typical value that is integrated from the autocorrelation function (ACF) to indicate whether the system converges. The documentation of \texttt{emcee} and \citet{emcee} suggests that $N > 50\tau$ is long enough, where $N$ is the length of the MCMC chain. We ran a chain of 2000 steps, and the autocorrelation time was $\tau=24.66$. We discarded the first $\lceil\tau+50\rceil$ steps, which may not converge well, and flattened the following steps to obtain a total of $1925\times512=985600$ samples.

\subsubsection{Results of the localized FRBs}

The histogram of all samples is plotted with a bin-width chosen by the Freedman Diaconis rule implemented by \texttt{numpy}. The probability density function (PDF) is given by the kernel density estimation (KDE). We also plot the cumulative histogram to obtain the cumulative distribution function (CDF). The best fit is $H_0=69.40_{-1.97}^{+2.14}$ \kmspc~, with a $1\sigma$ confidence interval, as shown in Fig.~\ref{fig:H_0_localized}. Our constraint from localized FRBs lies between the early-time result given by \citet{Intro-Planck2018} and the late-time result given by \citet{Intro-SNIa}.

Moreover, we tried to divide the FRBs into several bins to investigate the value of $H_0$ at different redshifts, similar as \citep{Krishnan2020,Jia2023,Intro-HubbleTension2,OColgain2024,Jia2025}. We divided the FRBs into two bins and determined an apparently descending trend, which is not significant. When we tried to divide the sample into more bins, the values became unstable and highly sensitive to the selection of a single FRB. This might be caused by the limited number and uneven distribution of the FRBs.

\begin{figure}
	\includegraphics[width=\columnwidth]{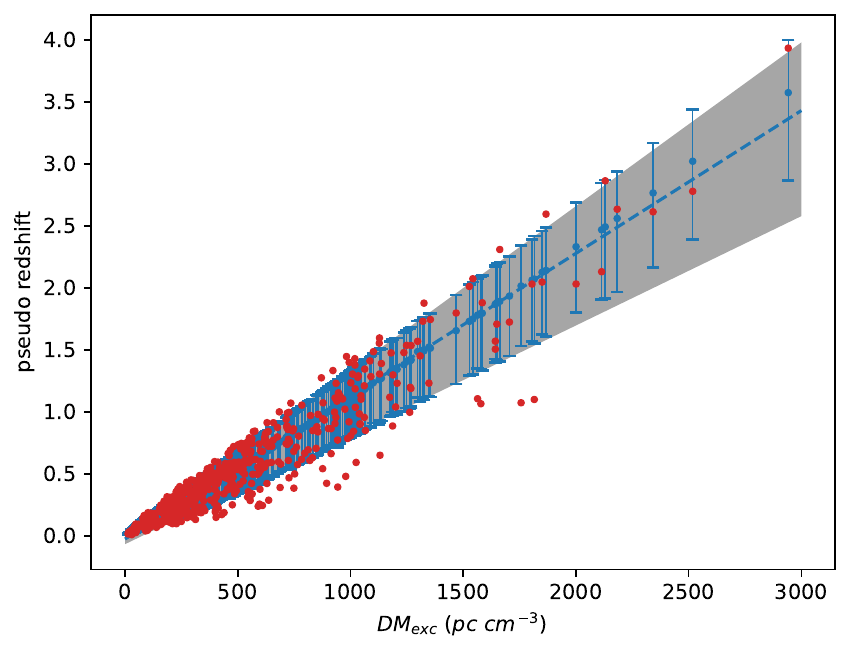}
    \centering
    \caption{Pseudo-redshifts of nonlocalized FRBs in the first CHIME catalog. The blue dots with the error bars are the median values and 1$\sigma$ confidence interval of redshifts estimated with the MCMC method. The dashed blue line and gray region were fitted from the blue dots. The red dots are the pseudo-redshifts (see Section~\ref{sec:scattered_redshifts}), which were decided randomly based on the PDF given by the MCMC sampling for each FRB. Extreme MCMC samples were neglected.}
    \label{fig:fixed_redshift_scatter}
\end{figure}
\begin{figure}
	\includegraphics[width=\columnwidth]{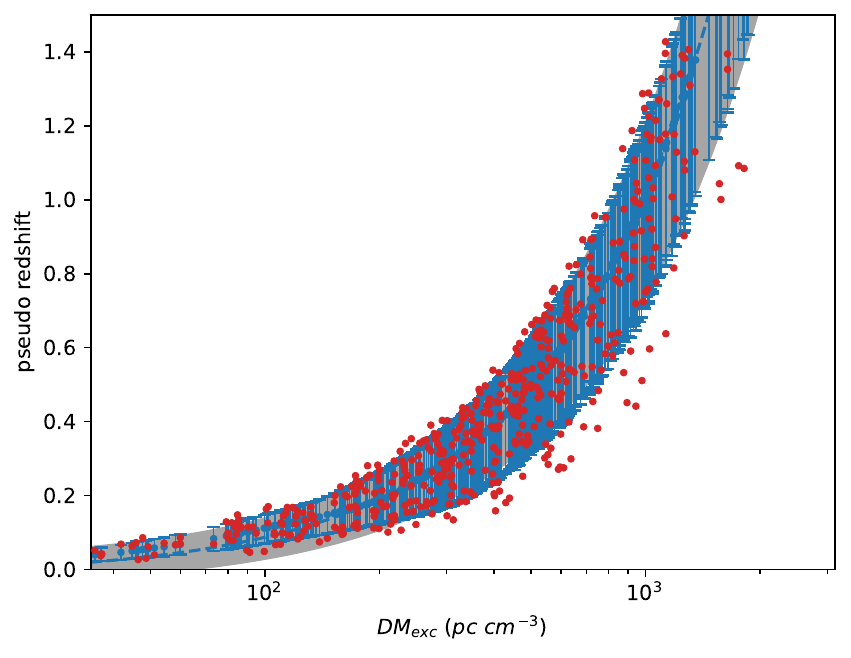}
    \centering
    \caption{Log scale plot of the pseudo-redshifts in Fig.~\ref{fig:fixed_redshift_scatter}}
    \label{fig:fixed_redshift_scatter_log}
\end{figure}

\section{Constraint with nonlocalized FRBs}

\label{sec:unlocalized}
Although the number of localized FRBs is increasing rapidly, most FRBs still remain nonlocalized. It is therefore crucial to use nonlocalized FRBs. A possible solution is to inverse the pseudo-redshifts with the observed DM \citep{Lin-unlocalized}. Compared with another method such as generating FRB data with simulation, FRBs with pseudo-redshifts are not dependent on any assumption of the DM distribution because we use real DM data as its foundation. We used all bursts in the first CHIME\footnote{The Canadian Hydrogen Intensity Mapping Experiment} catalog \citep{CHIME_catalog_1} and part of the available repeating bursts with definite coordinates in the CHIME catalog 2023 \citep{CHIME_Catalog_2023} to run the MCMC sampling. When we processed the repeating FRBs, we considered all bursts from the same source as one single eventd and calculate their mean DM as \DMobs. With the pseudo-redshifts, we used all nonlocalized FRBs as localized FRBs to constrain $H_0$.

\subsection{Redshift distribution}
\label{sec:pseudo_redshift}

\subsubsection{Circular argument}
Before the pseudo-redshifts are estimated, an assumed value for $H_0$ is required. This is a circular argument. However, it can be considered as an iterative analysis similar to the Newton-Raphson method. We assumed an initial value $H_0^{(0)}$ to calculate the pseudo-redshifts $z^{(0)}$ in the first iteration, and we applied $z^{(0)}$ to estimate $H_0^{(1)}$. To be precise, this step should be repeated as
\begin{equation}
H_0^{(0)}\rightarrow z^{(0)}\rightarrow H_0^{(1)}\rightarrow z^{(1)}\rightarrow \cdots\rightarrow H_0^{(n)}\rightarrow z^{(n)}
\label{eq:circular}
\end{equation}
until $\lvert H_0^{(n)}-H_0^{(n+1)}\rvert<\varepsilon$. Through the MCMC sampling, we found that different initial values of $H_0$ affect the pseudo-redshifts and the final estimation of $H_0$ only little.

Another way to avoid a circular argument is to consider $H_0$ as an unfit parameter (same as the pseudo-redshift) instead of assuming its value. To guarantee that the estimated $H_0$ is the same for all FRBs, the pseudo-redshifts must be simultaneously computed for all FRBs, that is, $(H_0,z_1,z_2,\cdots,z_n)$ must be fit simultaneously. For $n=527$, an enormous computational resource is required to fit a 528-dimensional parameter space. We used the first method.

\subsubsection{Calculating the redshift distribution}

The pseudo-redshifts can be estimated with two different methods: the maximum likelihood estimation (MLE), and a Monte Carlo sampling. Both methods require a likelihood function that is slightly different from Equation~(\ref{eq:likelihood_FRB}). $H_0$ is a known parameter, and $z_i$ is the parameter we need to fit. The equation is rewritten as
\begin{equation}
\begin{aligned}
\mathcal{L}(z_i) = &\int_0^{(1+z_i){({\rm DM}_i-{\rm DM_{MW}})}} p_{\rm host}({\rm DM_{host}} \mid z_i)\ \times\\
&p_{\rm IGM}({{\rm DM}_i-{\rm DM_{MW}} - \frac{\rm DM_{host}}{(1+z_i)}}\mid z_i)\ d\ {\rm DM_{host}}.
\label{eq:likelihood_pseudo}
\end{aligned}
\end{equation}
The MLE can give the mean value of pseudo-redshift for each FRB. The chain may fail to converge for FRBs with low \DMexc, and MLE gives no information about its distribution. The Monte Carlo sampling provides the probability density distribution and works for FRBs with a low DM. As a result, we used a Monte Carlo sampling instead of the MLE.

With a similar MCMC sampling as described in Section~\ref{sec:MCMC_local}, 57600 samples ($n_{\rm walkers}$=64, discard=100, steps=1000) were generated for each FRB. We calculated the 16, 50, and 84 percentiles (1$\sigma$ confidence) of the pseudo-redshift for each FRB and plot them in the form of error bars on a scatter plot with the DM as the horizontal axis. For any given percentile (e.g., $z_{16},z_{50}$, or $z_{84}$), $z\sim{\rm DM_{exc}}$ shows a good linear relation that agrees with the pseudo-redshifts from previous research. We plot the result of the linear regression and show the 68\% confidence region in Fig.~\ref{fig:fixed_redshift_scatter} and Fig.~\ref{fig:fixed_redshift_scatter_log}.

\subsection{MC sampling of nonlocalized FRBs}
\subsubsection{Generating the pseudo-redshifts}
\label{sec:scattered_redshifts}
The pseudo-redshifts have an intrinsic difference from real redshifts of FRBs. The errors of the real redshifts are negligible. However, the distribution of the pseudo-redshifts could not be constrained within a small confidence interval. It is insufficient to simply use an error bar or Gaussian distribution around the peak value to describe pseudo-redshifts.

We assigned a pseudo-redshift within a relatively large interval for each FRB based on its PDF. It is located near the peak with a high probability and lies far away from the peak with a very low probability.
When we repetitively generated this hundreds of times, all these pseudo-redshifts reflected the PDF well. Although we were unable to generate multiple times for one FRB during the sampling, we were able to generate once for each FRB, and we had 527 FRBs. It is statistically safe to reflect their PDFs, and this is exactly how real redshifts are like. Furthermore, we did not consider an error bar of the pseudo-redshift because an error bar is insufficient to describe the PDF.

There is a small possibility that the generated redshift is very far away from the peak. To avoid this extreme situation, we discarded generated redshifts that lay outside 94\% of the region that was centered the peak (i.e., below 3\% and above 97\% in the PDF) and regenerate again. The pseudo-redshifts are shown as red dots in Fig.~\ref{fig:fixed_redshift_scatter} and Fig.~\ref{fig:fixed_redshift_scatter_log}.

\subsubsection{$H_0$ from nonlocalized FRBs}

\begin{figure}
	\includegraphics[width=\columnwidth]{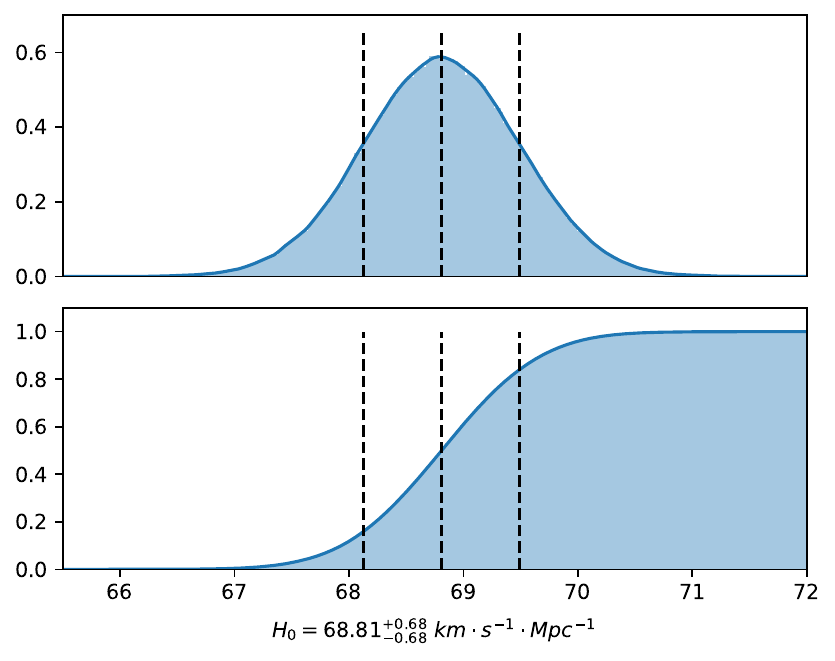}
    \centering
    \caption{PDF and CDF of the $H_0$ given by the pseudo-redshifts of nonlocalized FRBs. The vertical line shows the result $H_0=68.81_{-0.68}^{+0.68}$ \kmspc with 1$\sigma$ uncertainty.}
    \label{fig:H_0_unlocal_fixed}
\end{figure}

We ran an MCMC sampling similar to Section~\ref{sec:localized} using the pseudo-redshifts as described in Section~\ref{sec:scattered_redshifts}. The result was $H_0=68.81_{-0.68}^{+0.68}$ \kmspc, as shown in Fig.~\ref{fig:H_0_unlocal_fixed}. The median value given by nonlocalized FRBs is very close to that given by localized FRBs. Moreover, since the redshifts are pseudo, the uncertainty of the result from the nonlocalized FRBs is much more inspiring than the median value. It provides a convincing prediction that if 527 FRBs are localized, the uncertainty drops to $\sim1\%$ at a confidence level of $1\sigma$. Compared with the result of localized FRBs, we have $err_1/err_2=3.02$ and $\sqrt{N_2}/\sqrt{N_1}=2.21$. The ratio is roughly consistent with the relation $err\sim1/\sqrt{N}$.

\section{Discussion}
\label{sec:discussion}
By using the pseudo-redshifts of 527 nonlocalized CHIME FRBs, we significantly reduced the uncertainty of the constraint. However, some bias and errors must be included for a full discussion. We generally divide them into the statistical and the systematic error.

\subsection{Statistical errors}

\begin{figure}
	\includegraphics[width=\columnwidth]{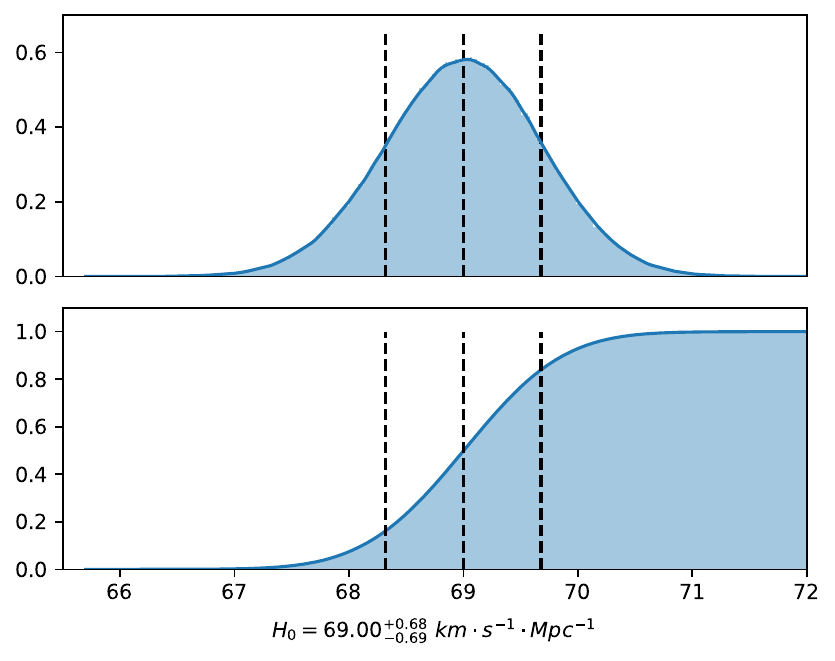}
    \centering
    \caption{PDF and CDF of $H_0$ considering the statistical error of $\Omega_m$ using the pseudo-redshifts of nonlocalized FRBs. The vertical line shows the result $H_0=69.00_{-0.69}^{+0.68}$ \kmspc with the 1$\sigma$ uncertainty.}
    \label{fig:H_0_statistic}
\end{figure}

The statistical error mainly refers to the error of the cosmological parameters $\Omega_bh^2$ and $\Omega_m$ in Equation~(\ref{eq:DM_IGM_mean_2}).  Other constants such as $G$ and $m_p$ are already measured with extremely high precision. \citet{Intro-Planck2018} gave $\Omega_bh^2 = 0.02242 \pm 0.00014$ and $\Omega_m = 0.3111 \pm 0.0056$. 

For $\Omega_bh^2$, it appears in Equation~(\ref{eq:DM_IGM_mean_2}) as a linear term with $H_0$. Assuming a Gaussian distribution $p(x)$ of about $\mu=0.02242$, we roughly estimated
\begin{equation}
H_0\sim\frac{\Omega_bh^2}{\langle{\rm DM_{IGM}}\rangle}\sim\int^\infty_{-\infty}\frac{x}{\langle{\rm DM_{IGM}}\rangle}p(x)dx\sim\frac{\mu}{\langle{\rm DM_{IGM}}\rangle},
\end{equation}
which means that if $\Omega_bh^2$ follows a symmetric distribution (e.g., a Gaussian distribution), it affects $H_0$ only weakly. Furthermore, the error of $\Omega_bh^2$ ($\sim0.6\%$) is much smaller than for other terms such as $\Omega_m$, and it is also smaller than the result of our constraint ($\Delta H_0/H_0\sim0.94\%$). As a result, it is safe to ignore the error of $\Omega_bh^2$.

For $\Omega_m$, which appears inside the integral in Equation~(\ref{eq:DM_IGM_mean_2}), the uncertainty is $\sim1.8\%$ and cannot be ignored. To marginalize $\Omega_m$, the best way is to add another level of integral, which would significantly increase the computation time. Alternatively, we considered replacing the integral with an expansion. We assumed a Gaussian distribution $p(x)$ with $\mu=0.3111$ and $\sigma=0.0056$. The integral can then be written as
\begin{equation}
\begin{aligned}
\label{eq:expansion_1}
I&=\int_{-\infty}^{\infty} p(x)\int_{0}^{Z} \frac{(1+z)}{\left[x(1+z)^{3} + 1 - x \right]^{1 / 2}}dzdx\\
&\simeq\int_{0}^{Z}\frac{(1+z)}{I_0}\int_{\mu-\sigma}^{\mu+\sigma} \frac{p(x) }{\left[x(1+z)^{3} + 1 - x \right]^{1 / 2}}dxdz\\
&\equiv\int_{0}^{Z}\frac{(1+z)}{I_0}\int_{\mu-\sigma}^{\mu+\sigma}f(x,z)dxdz,
\end{aligned}
\end{equation}
where $I_0=\int_{\mu-\sigma}^{\mu+\sigma}p(x)dx=0.683$ is the normalization factor and $f(x,z)=p(x)/\left[x(1+z)^{3} + 1 - x \right]^{1 / 2}$ is our target function. $Z$ is the redshift of an FRB, and $z$ is our integration variable. Since both $\int f(x,z)dx$ and $\int f(x,z)dz$ cannot be expressed by elementary functions, we performed a series expansion on $f(x,z)dz$ around $x=\mu$ and obtained $g(x,z)=\sum_{i=0}^5a_i(z)(x-\mu)^i$. We plot the figure of $g(x,z)$ at different $z$ and compare it with the original function to ensure that the expansion is acceptable. The inner integral in Equation~(\ref{eq:expansion_1}) can then be written as an explicit function of $z$, that is, $h(z)=\int_{\mu-\sigma}^{\mu+\sigma}g(x,z)dx$. However, the form of $h(z)$ is still too complicated for an integration, and we thus needed to perform another series expansion around $z=z_0$. To determine the best value for $z_0$, we plot the expanded function at different $z_0$ and degrees. An expansion of $h(z)$ to the term of $(z-z_0)^4$ around $z_0=2.5$ provides best fit for both $z\rightarrow0$ and $z\rightarrow4$. We denote it as $j(z)=\sum_{i=0}^4b_i(z)(z-z_0)^i$, and we can complete the whole integral: $I\simeq1.02\ Z + 0.19\ Z^2 - 0.14\ Z^3 + 0.043\ Z^4 - 0.0066\ Z^5 + 0.00042\ Z^6$.

Taking the new expression into Equation~(\ref{eq:DM_IGM_mean_2}), we ran the MCMC sampling with $\Omega_m$ marginalized. The result is shown in Fig.~\ref{fig:H_0_statistic} (the method in Section~\ref{sec:scattered_redshifts} was used). $H_0$ is $69.00_{-0.69}^{+0.68}$ \kmspc in $1\sigma$ confidence intervals, which is consistent with the previous result in Fig.~\ref{fig:H_0_unlocal_fixed}. Therefore, the statistical error of $\Omega_m$ does not influence the constraint on $H_0$ significantly.

\subsection{Systematic errors}

\begin{figure}
	\includegraphics[width=\columnwidth]{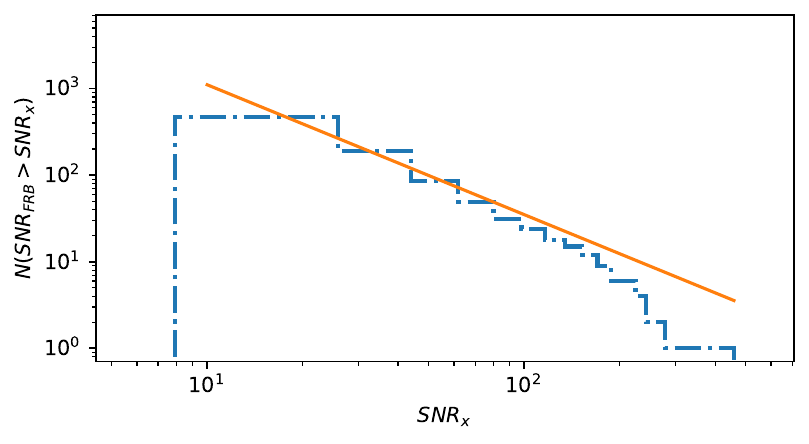}
    \centering
    \caption{Signal-to-noise ratio for one-off FRBs in the first CHIME catalog where repeating FRBs are not counted. The yellow line with a slope of -1.5 (in log-log scale) was not fit by the data and is shown just for comparison.}
    \label{fig:snr}
\end{figure}

\label{sec:systematic_error}
Several systematic errors need to be discussed. In Equation~(\ref{eq:DM_IGM_mean_2}), compared with $\Omega_bh^2$ and $\Omega_m$, the fraction of baryons in IGM $f_{\rm IGM}$ may introduce more uncertainty. However, we still know little about $f_{\rm IGM}$. \citet{fIGM} gave a value of $\sim0.84$. \citet{connor2024} provided a more accurate constraint of $\sim0.93$, which depends on the models. \citet{Khrykin2024ApJ} gave a lower value of $f_{\rm IGM}=0.59^{+0.11}_{-0.10}$ based on the FLIMFLAM spectroscopic survey. Current research is clearly still undetermined. Furthermore, it is difficult to separate the error of $f_{\rm IGM}$ from $H_0$ as they appear in a coupling term $f_{\rm IGM}/H_0$ in Equation~(\ref{eq:DM_IGM_mean_2}). To be precise, only a constraint on $f_{\rm IGM}/H_0$ can be made instead of a constraint on $H_0$. The value of $H_0$ must therefore be determined by other approaches when $f_{\rm IGM}$ is constrained from FRBs. By fixing $H_0$, it has been found that the uncertainty of $f_{\rm IGM}$ is about 8\% \citep{Yang2022ApJL,connor2024}. The systematic error from $f_{\rm IGM}$ clearly currently dominates the error of the measured $H_0$. On the other hand, it may vary with redshift. Without further independent constraints on $f_{\rm IGM}$, we cannot exclude the error of $f_{\rm IGM}$. Trying to marginalize $f_{\rm IGM}$ with a Gaussian distribution during the MCMC sampling was unable to provide a more accurate value of $H_0$. It only gives $\Delta H_0/H_0\sim\sigma_{f_{\rm IGM}}$ where $\sigma_{f_{\rm IGM}}$ is the error assumed in the Gaussian distribution. More research with other methods is required to further investigate the error of $f_{\rm IGM}$.

\DMhost ~includes all contributions to the DM that come from the host galaxy. Based on IllustrisTNG simulations, the probability distribution of \DMhost including the local cosmic structure (e.g., filament) halo and interstellar medium of the host were derived \citep{Zhang2020host}. However, the vicinity of FRB progenitors was not considered. The most promising progenitors of FRBs are young magnetars \citep{Wang2017}, which can be formed by the core collapse of massive stars or by mergers of two compact objects \citep{Wang2020}. They might therefore be embedded in a magnetar wind nebula and supernova remnant \citep{Yang2017,Piro2018,Zhao2021}. On the other hand, the large \DMhost with a rotation measure reversal for FRB 20190520B indicates that it might reside in a binary system \citep{Wang2022,Anna-Thomas2023}. Similar FRBs should therefore be removed when measuring $H_0$. For FRBs with little DM contribution from the vicinity, a precise modeling should be performed. It is important to use optical observations of the FRB host galaxy environment, combined with the rotation measure and the scattering times of FRBs to constrain \DMhost ~\citep{Cordes2022}.

\DMIGM ~refers to the entire DM contribution between the host galaxy and the Milky Way, which includes contributions from the IGM and other intervening galaxy groups and halos. It is difficult to determine the number of intervening galaxies because the randomness is significant, and it is even more difficult to identify the stellar mass and SFR for each galaxy, which is greatly relevant to the relating DM. \citet{connor2024} tried to separate the contribution of the IGM and halos and reported a corresponding baryon fraction. Because we chose their value of $f_{\rm IGM}$, it contains the baryon from the IGM and halos, which can be considered as a statistical average to compensate for the unknown intervening galaxies.

\begin{figure}
	\includegraphics[width=\columnwidth]{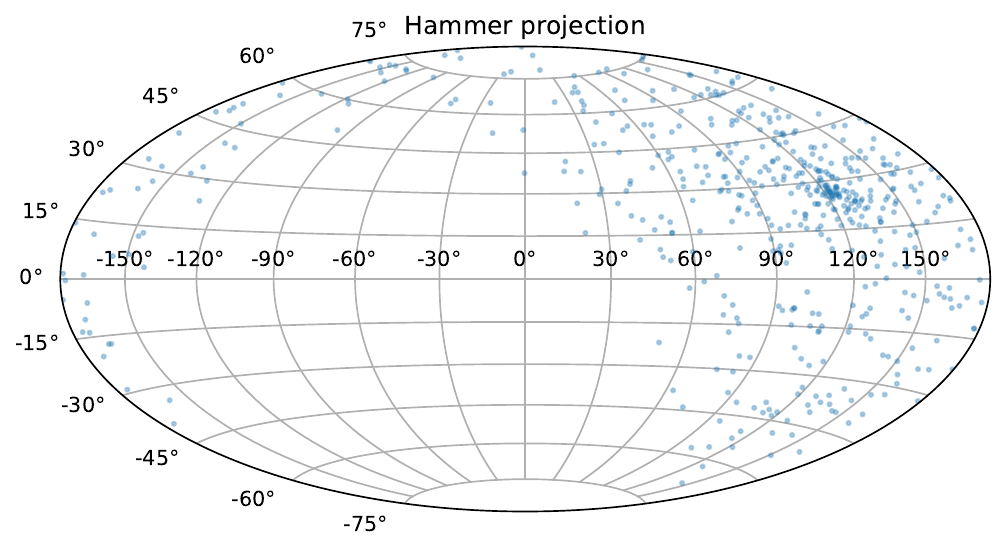}
    \centering
    \caption{Hammer projection of the FRB distribution in the first CHIME catalog  in the galactic coordinate system. Repeating FRBs are only counted once.}
    \label{fig:galactic_hammer}
\end{figure}

Several selection biases should be discussed, such as the signal-to-noise ratio (S/N) effect, the ISM effect, the selection effect of nonlocalized FRBs, and the gridding effect \citep{James2022}. For the selection effect of nonlocalized FRBs, our constraint has no such error because we made use of most nonlocalized FRBs in the CHIME database. For the gridding effect, we used continuous values for redshifts and DMs instead of discrete variables. For the S/N effect, the log-log figure of a number of events observed above the S/N threshold should follow a power law of -1.5 (in the log-log plot, i.e., $N\propto {\rm S/N}^{-1.5}$), and \citet{James2022} found that events from CRAFT/ICS deviate from the -1.5 power law. We plot the same figure with nonlocalized FRBs from the CHIME database in Fig.~\ref{fig:snr}. The histogram followed the power law well, and our constraint is therefore not much influenced by the S/N effect. For the ISM effect, \citet{James2022} claimed that \DMISM would increase at low galactic latitudes, which may prevent telescopes from observing these events. We plot the Hammer projection of FRBs in the galactic coordinate system in Fig.~\ref{fig:galactic_hammer}. A Hammer projection is an equal-area projection, and a considerable number of FRBs are located in low galactic latitude areas. Furthermore, a few of these low-galactic-latitude FRBs have shown high values of \DMISM. Only one of the 108 localized FRBs has \DMISM$>200$ \pccm, but the largest observed DM is about 1500 \pccm. It is not likely that FRBs are missed by observations because \DMISM is too high.

\section{Conclusions}
We ran an MCMC sampling to constrain $H_0$ using 108 localized FRBs and 527 nonlocalized FRBs from the CHIME catalog. We applied the redshift-DM relation and a Bayesian estimation to build the MCMC model. We used normalization factors obtained from the IllustrisTNG simulation to model the DM distribution. For localized FRBs, we obtained $H_0=69.40_{-1.97}^{+2.14}$ \kmspc with 108 FRBs, which lies between constraints from late- and early-time research. For nonlocalized FRBs, we ran individual MCMC samplings instead of a maximum likelihood estimation to obtain a probability density distribution of the pseudo-redshift for each FRB. We assigned pseudo-redshifts for FRBs and obtained $H_0=68.81_{-0.68}^{+0.68}$ \kmspc.

The statistical errors of the cosmological parameters, the systematic error of $f_{\rm IGM}$, and the selection biases were discussed. We showed that the statistical error of $\Omega_b$ affects our constraint less than $\Omega_m$. We performed a series expansion to marginalize $\Omega_m$ and obtained $H_0=69.00_{-0.69}^{+0.68}$ \kmspc. The degeneracy effect prevented us from separating the error of $f_{\rm IGM}$ from $H_0$. The uncertainty of $H_0$ is dominated by the error of the fraction of cosmic baryons in the diffuse ionized gas $f_{\rm IGM}$. Other systematic errors were neglected.

Our study predicts future constraints on $H_0$ with more localized FRBs. Our result shows that the uncertainty of $H_0$ is likely to drop to $\sim1\%$ when the number of localized FRBs increases to $\sim500$. FRBs will become a powerful tool for solving the Hubble tension. 

\begin{acknowledgements}
      We thank the anonymous referee for constructive comments. This work was supported by the National Natural Science Foundation of China (grant Nos. 12494575, 12273009 and 12393812), the National SKA Program of China (grant Nos. 2020SKA0120302 and 2022SKA0130100), and the Natural Science Foundation of Xinjiang Uygur Autonomous Region (grant No. 2023D01E20).
\end{acknowledgements}

\clearpage
\bibliographystyle{aa}
\bibliography{ms}

\begin{appendix}
\onecolumn
\section{FRB data}
\begin{table*}[h!]
	\centering
    \renewcommand\arraystretch{1.3}
   
	\caption{Properties of localized FRBs.}
    
	\label{tab:localized_FRBs}
        \begin{tabular}{ccccccc}
            \hline
            \hline
            
            FRB Type  & TNS Name & RA    & DEC   & DM ($\rm pc\ cm^{-3}$) & Redshift & Reference \\
            \hline
            \multirow{2}[1]{*}{1} & \multirow{2}[1]{*}{FRB20121102} & \multirow{2}[1]{*}{5:31:58} & \multirow{2}[1]{*}{+33:08:04} & \multirow{2}[1]{*}{557} & \multirow{2}[1]{*}{0.1927} & \citet{Tendulkar2016}\\
            &       &       &       &       &       & \citet{Chatterjee2017} \\
            3& FRB20171020A& 22:15:24.75& -19:35:07.00& 114.1& 0.0087& \citet{Mahony2018}\\
            1& FRB20180301& 6:12:54.44& +4:40:15.8& 536& 0.3304& \citet{Bhandari2022}\\
            2& FRB20180814& 4:22:56.01& +73:39:40.7& 189.4& 0.068& \citet{Michilli2023}\\
            2& FRB20180916& 1:58:00.75& +65:43:00.32& 348.76& 0.0337& \citet{Marcote2020}\\
            3& FRB20180924& 21:44:25.3& -40:54:00.1& 361.42& 0.3214& \citet{Bannister2019}\\
            1& FRB20181030*& 10:34:20.1& +73:45:05& 103.5& 0.0039& \citet{Bhardwaj2021}\\
            3& FRB20181112& 21:49:23.63& -52:58:15.39& 589.27& 0.4755& \citet{Prochaska2019}\\
            3& FRB20181220A& 23:14:52& +48:20:25& 209.4& 0.02746& \citet{Bhardwaj2024}\\
            3& FRB20181223C& 12:03:43& +27:33:09& 112.5& 0.03024& \citet{Bhardwaj2024}\\
            3& FRB20190102& 21:29:39.76& -79:28:32.5& 364.5& 0.2913& \citet{Bhandari2020}\\
            2& FRB20190110C& 16:37:16.43& +41:26:36.30& 221.6& 0.12244& \citet{Ibik2024}\\
            1& FRB20190303A& 13:51:58& 	+48:7:20& 222.4& 0.064& \citet{Michilli2023}\\
            3& FRB20190418A& 04:23:16& +16:04:02& 184.5& 0.07132& \citet{Bhardwaj2024}\\
            3& FRB20190425A& 17:02:42& +21:34:35& 128.2& 0.03122& \citet{Bhardwaj2024}\\
            1& FRB20190520B*& 16:02:04.266& -11:17:17.33& 1210.3& 0.241& \citet{Niu2022}\\
            3& FRB20190523& 13:48:15.6& +72:28:11& 760.8& 0.66& \citet{Ravi2019}\\
            3& FRB20190608& 22:16:04.74& -7:53:53.6& 339.5& 0.11778& \citet{Chittidi2021}\\
            3& FRB20190611& 21:22:58.91& -79:23:51.3& 321.4& 0.378& \citet{Heintz2020}\\
            3& FRB20190614& 4:20:18.13& +73:42:22.9& 959.2& 0.6& \citet{Law2020}\\
            1& FRB20190711& 57:40.7& -80:21:28.8& 593.1& 0.522& \citet{Heintz2020}\\
            3& FRB20190714& 12:15:55.12& -13:01:15.7& 504.13& 0.2365& \citet{Heintz2020}\\
            3& FRB20191001& 21:33:24.373& -54:44:51.43& 507.9& 0.234& \citet{Heintz2020}\\
            1& FRB20191106C& 13:18:19.23& +42:59:58.97& 332.2& 0.10775& \citet{Ibik2024}\\
            3& FRB20191228& 22:57:43.3& -29:35:38.7& 297.5& 0.2432& \citet{Bhandari2022}\\
            3& FRB20200120E*& 9:57:54.7& +68:49:0.9& 87.8& 0.0008& \citet{Kirsten2022}\\
            2& FRB20200223B& 00:33:04.68& +28:49:52.60& 201.8& 0.06024& \citet{Ibik2024}\\
            3& FRB20200430& 15:18:49.54& +12:22:36.8& 380.25& 0.16& \citet{Heintz2020}\\
            3& FRB20200906& 3:3:59.08& -14:04:59.5& 577.8& 0.3688& \citet{Bhandari2022}\\
            3& FRB20201123A& 17:34:40.8& -50:40:12& 433.55& 0.0507& \citet{Rajwade2022}\\
            2& FRB20201124& 5:08:03.5& +26:03:38.4& 413.52& 0.098& \citet{Ravi2022}\\
            3& FRB20210117A& 22:39:55.015& -16:09:05.45& 728.95& 0.214& \citet{Bhandari2023}\\
            3& FRB20210320C& 13:37:50.10& -16:07:21.6& 384.8& 0.2797& \citet{Shannon2024}\\
            3& FRB20210405I*& 17:01:21.5& -49:32:42.5& 565.17& 0.066& \citet{Driessen2023}\\
            3& FRB20210410D& 21:44:20.7& -79:19:05.5& 578.78& 0.1415& \citet{Caleb2023}\\
            3& FRB20210603A& 0:41:05.774& +21:13:34.573& 500.147& 0.177& \citet{Cassanelli2023}\\
            3& FRB20210807D& 19:56:53.14& -00:45:44.50& 251.3& 0.1293& Deller(in prep.)\\
            3& FRB20211127I& -& -& 234.83& 0.0469& Deller(in prep.)\\
            3& FRB20211203C& 13:38:15.00& -31:22:48.20& 635& 0.3439& \citet{Gordon2023}\\
            \hline
        \end{tabular}
        \tablefoot{FRB type shows whether an event is an one-off burst (type = 3) or a repeated burst (type = 1 or 2). Type = 1 indicates a 20121102-like burst and type = 2 indicates a 20180916-like burst. FRB types are required while using the \DMhost ~distributions given by IllustrisTNG simulation.  FRB20190520B and FRB20220831A have extreme $\rm DM_{host}$. FRB20181030, FRB20200120E, FRB20220319D and FRB20210405I are excluded due to ${\rm DM_{exc}<0}$. FRB20221027A is excluded for its ambiguity in host galaxy localization.}
\end{table*}

\begin{table*}[h!]
	\centering
        \renewcommand\arraystretch{1.3}
	\caption*{(continued)}
        \begin{tabular}{ccccccc}
            \hline
            \hline
            FRB Type  & TNS Name & RA    & DEC   & DM ($\rm pc\ cm^{-3}$) & Redshift & Reference \\
            \hline
            3& FRB20211212A&  -& -& 206& 0.0715& Deller(in prep.)\\
			3& FRB20220105A& 13:55:12.94& +22:27:59.40& 580& 0.2785& \citet{Gordon2023}\\
            3& FRB20220204A& 18:16:54.30& +69:43:21.01& 612.2& 0.4& \citet{Sharma2024}\\ 
            3& FRB20220207C& 20:40:47.886& +72:52:56.378& 262.38& 0.04304& \citet{Law2024}\\ 
            3& FRB20220208A& 21:30:18.03& +70:02:27.75& 437& 0.351& \citet{Sharma2024}\\ 
            3& FRB20220307B& 23:23:29.88& +72:11:32.6& 499.27& 0.248123& \citet{Law2024}\\ 
            3& FRB20220310F& 8:58:52.9& +73:29:27.0& 462.24& 0.477958& \citet{Law2024}\\ 
            3& FRB20220319D*& 08:42.7& +71:02:06.9& 110.95& 0.0111& \citet{Ravi2023-2}\\ 
            3& FRB20220330D& 10:55:00.30& +70:21:02.70& 468.1& 0.3714& \citet{Sharma2024}\\ 
            3& FRB20220418A& 14:36:25.34& +70:05:45.4& 623.25& 0.622& \citet{Law2024}\\ 
            3& FRB20220501C& 23:29:31.00& -32:29:26.6& 449.5& 0.381& \citet{Shannon2024}\\ 
            3& FRB20220506D& 21:12:10.76& +72:49:38.2& 396.97& 0.30039& \citet{Law2024}\\ 
            3& FRB20220509G& 18:50:40.8& +70:14:37.8& 269.53& 0.0894& \citet{Law2024}\\ 
            1& FRB20220529A& 01:16:25.01& +20:37:57.03& 246& 0.1839& Li(in prep.)\\ 
            3& FRB20220610A& 23:24:17.569& -33:30:49.37& 1457.624& 1.016& \citet{Ryder2023}\\ 
            3& FRB20220717A& 19:33:13.0& -19:17:15.8& 637& 0.36295& \citet{Rajwade2024}\\ 
            3& FRB20220725A& 23:33:15.65& -35:59:24.9& 290.4& 0.1926& \citet{Shannon2024}\\ 
            3& FRB20220726A& 04:55:46.96& +69:55:44.80& 686.55& 0.361& \citet{Sharma2024}\\ 
            3& FRB20220825A& 20:47:55.55& +72:35:05.9& 651.24& 0.241397& \citet{Law2024}\\ 
            3& FRB20220831A*& 22:34:46.93& +70:13:56.50& 1146.25& 0.262& \citet{connor2024}\\ 
            2& FRB20220912A& 23:09:04.9& +48:42:25.4& 219.46& 0.0771& \citet{Ravi2023-1}\\ 
            3& FRB20220914A& 18:48:13.63& +73:20:12.9& 631.28& 0.1139& \citet{Law2024}\\ 
            3& FRB20220918A& 01:10:22.11& -70:48:41.0& 656.8& 0.491& \citet{Shannon2024}\\ 
            3& FRB20220920A& 16:01:01.70& +70:55:07.7& 314.99& 0.158239& \citet{Law2024}\\ 
            3& FRB20221012A& 18:43:11.69& +70:31:27.2& 441.08& 0.284669& \citet{Law2024}\\ 
            3& FRB20221027A*& 08:43:29.23& +72:06:03.50& 452.5& 0.229& \citet{Sharma2024}\\ 
            3& FRB20221029A& 09:27:51.22& +72:27:08.34& 1391.05& 0.975& \citet{Sharma2024}\\ 
            3& FRB20221101B& 22:48:51.89& +70:40:52.20& 490.7& 0.2395& \citet{Sharma2024}\\ 
            3& FRB20221106A& 03:46:49.15& -25:34:11.3& 343.8& 0.2044& \citet{Shannon2024}\\ 
            3& FRB20221113A& 04:45:38.64& +70:18:26.60& 411.4& 0.2505& \citet{Sharma2024}\\ 
            3& FRB20221116A& 01:24:50.45& +72:39:14.10& 640.6& 0.2764& \citet{Sharma2024}\\ 
            3& FRB20221219A& 17:10:31.15& +71:37:36.63& 706.7& 0.554& \citet{Sharma2024}\\ 
            3& FRB20230124& 15:27:39.90& +70:58:05.20& 590.6& 0.094& \citet{Sharma2024}\\ 
            3& FRB20230203A& 10:06:38.7816& +35:41:38.76& 420.1& 0.1464& \citet{Amiri2025}\\ 
            3& FRB20230216A& 10:25:53.32& +03:26:12.57& 828& 0.531& \citet{Sharma2024}\\ 
            3& FRB20230222A& 07:07:50.4864& +11:13:28.272& 706.1& 0.1223& \citet{Amiri2025}\\ 
            3& FRB20230222B& 15:54:57.3888& +30:53:55.32& 187.8& 0.11& \citet{Amiri2025}\\ 
            3& FRB20230307A& 11:51:07.52& +71:41:44.30& 608.9& 0.271& \citet{Sharma2024}\\ 
            3& FRB20230311A& 06:04:26.3184& +55:56:45.42& 364.3& 0.1918& \citet{Amiri2025}\\ 
            3& FRB20230501A& 22:40:06.52& +70:55:19.82& 532.5& 0.301& \citet{Sharma2024}\\ 
            1& FRB20230506C& 00:48:23.9121& +42:00:21.954& 766.5& 0.3896& \citet{AnnaThomas2025}\\ 
            3& FRB20230521B& 23:24:08.64& +71:08:16.91& 1342.9& 1.354& \citet{connor2024}\\ 
            3& FRB20230526A& 01:28:55.83& -52:43:02.4& 361.4& 0.157& \citet{Shannon2024}\\ 
            3& FRB20230626A& 15:42:31.10& +71:08:00.77& 451.2& 0.327& \citet{Sharma2024}\\ 
            3& FRB20230628A& 11:07:08.81& +72:16:54.64& 345.15& 0.1265& \citet{Sharma2024}\\ 
        \hline
        \end{tabular}
\end{table*}

\begin{table*}[h!]
	\centering
        \renewcommand\arraystretch{1.3}
	\caption*{(continued)}
        \begin{tabular}{ccccccc}
            \hline
            \hline
            FRB Type  & TNS Name & RA    & DEC   & DM ($\rm pc\ cm^{-3}$) & Redshift & Reference \\
            \hline
            3& FRB20230703A& 12:18:29.868& +48:43:47.748& 291.3& 0.1184& \citet{Amiri2025}\\ 
            3& FRB20230708A& 20:12:27.73& -55:21:22.6& 411.51& 0.105& \citet{Shannon2024}\\ 
            3& FRB20230712A& 11:09:26.05& +72:33:28.02& 586.96& 0.4525& \citet{Sharma2024}\\ 
            3& FRB20230718A& 08:32:38.86& -40:27:07.0& 477& 0.035& \citet{Shannon2024}\\ 
            3& FRB20230730A& 03:38:39.4944& +33:09:33.48& 312.5& 0.2115& \citet{Amiri2025}\\ 
            3& FRB20230814A& 22:23:53.94& +73:01:33.26& 696.4& 0.5535& \citet{connor2024}\\ 
            3& FRB20230902A& 03:28:33.55& -47:20:00.6& 440.1& 0.3619& \citet{Shannon2024}\\ 
            3& FRB20230926A& 17:56:29.9712& +41:48:51.48& 222.8& 0.0553& \citet{Amiri2025}\\ 
            3& FRB20230930A& 00:42:01.676& +41:25:3.143& 456& 0.0925& \citet{AnnaThomas2025}\\ 
            3& FRB20231005A& 16:24:06.72& +35:26:55.356& 189.4& 0.0713& \citet{Amiri2025}\\ 
            3& FRB20231011A& 01:12:57.864& +41:44:56.76& 186.3& 0.0783& \citet{Amiri2025}\\ 
            3& FRB20231017A& 23:07:01.0296& +36:39:09.648& 344.2& 0.245& \citet{Amiri2025}\\ 
            3& FRB20231025B& 18:03:09.1368& +63:59:20.688& 368.7& 0.3238& \citet{Amiri2025}\\ 
            3& FRB20231120A& 09:35:56.15& +73:17:04.80& 438.9& 0.07& \citet{Sharma2024}\\ 
            3& FRB20231123A& 05:30:29.58& +04:28:31.944& 302.1& 0.0729& \citet{Amiri2025}\\ 
            3& FRB20231123B& 16:10:09.16& +70:47:06.20& 396.7& 0.2625& \citet{Sharma2024}\\ 
            3& FRB20231201A& 03:38:21.4296& +26:49:03.612& 169.4& 0.1119& \citet{Amiri2025}\\ 
            3& FRB20231206A& 07:29:46.2816& +56:15:22.572& 457.7& 0.0659& \citet{Amiri2025}\\ 
            3& FRB20231220A& 08:15:38.09& +73:39:35.70& 491.2& 0.3355& \citet{connor2024}\\ 
            3& FRB20231223C& 17:18:10.716& +29:29:52.584& 165.8& 0.1059& \citet{Amiri2025}\\ 
            3& FRB20231226A& 10:21:27.30& +06:06:36.9& 329.9& 0.1569& \citet{Shannon2024}\\ 
            3& FRB20231229A& 01:45:52.2792& +35:06:46.512& 198.5& 0.019& \citet{Amiri2025}\\
            1& FRB20240114A& 21:27:39.835& +4:19:45.634& 527.7& 0.13& Chen(in prep.)\\ 
            3& FRB20240119A& 14:57:52.12& +71:36:42.33& 483.1& 0.37& \citet{connor2024}\\ 
            3& FRB20240123A& 04:33:03.00& +71:56:43.02& 1462& 0.968& \citet{connor2024}\\ 
            3& FRB20240201A& 09:59:37.34& +14:05:16.9& 374.5& 0.042729& \citet{Shannon2024}\\ 
            3& FRB20240210A& 00:35:07.10& -28:16:14.7& 283.73& 0.023686& \citet{Shannon2024}\\ 
            3& FRB20240213A& 11:04:40.39& +74:04:31.40& 357.4& 0.1185& \citet{connor2024}\\ 
            3& FRB20240215A& 17:53:45.90& +70:13:56.50& 549.5& 0.21& \citet{connor2024}\\ 
            3& FRB20240229A& 11:19:56.05& +70:40:34.40& 491.15& 0.287& \citet{connor2024}\\ 
            3& FRB20240310A& 01:10:29.25& -44:26:21.9& 601.8& 0.127& \citet{Shannon2024}\\ 
        \hline
        \end{tabular}
\end{table*}
\FloatBarrier
\twocolumn
\end{appendix}

\end{document}